\documentclass{ifacconf}

\usepackage{graphicx}      
\usepackage{natbib}        
\usepackage{amsfonts,amsmath}
\usepackage{subfig}   
\usepackage{pict2e}  	
\usepackage[svgnames]{xcolor}   

\newtheorem{remark}{Remark}
\newtheorem{example}{Example}
\DeclareRobustCommand{\bbone}{\text{\usefont{U}{bbold}{m}{n}1}}  

\newcommand{\R}{\mathbb R}

\begin{document}
\begin{frontmatter}

\title{Understanding Human Mobility Flows from Aggregated Mobile Phone Data\thanksref{footnoteinfo}} 
\thanks[footnoteinfo]{This work was supported by funding from project MIE - Mobilit\`a Intelligente Ecosostenibile (CTN01\_00034\_594122), Cluster ``Tecnologie per le Smart Communities''.}

\author[First]{Caterina Balzotti}
\author[Second]{Andrea Bragagnini}
\author[Third]{Maya Briani} 
\author[Fourth]{Emiliano Cristiani} 

\address[First]{Istituto per le Applicazioni del Calcolo, Consiglio Nazionale delle Ricerche, Rome, Italy (c.balzotti@iac.cnr.it)}
\address[Second]{TIM Services Innovation, Italy (andrea.bragagnini@telecomitalia.it)}
\address[Third]{Istituto per le Applicazioni del Calcolo, Consiglio Nazionale delle Ricerche, Rome, Italy (m.briani@iac.cnr.it)}
\address[Fourth]{Istituto per le Applicazioni del Calcolo, Consiglio Nazionale delle Ricerche, Rome, Italy (e.cristiani@iac.cnr.it)}

\begin{abstract}                
In this paper we deal with the study of travel flows and patterns of people in large populated areas. 
Information about the movements of people is extracted from coarse-grained aggregated cellular network data without tracking mobile devices individually. 
Mobile phone data are provided by the Italian telecommunication company TIM and consist of density profiles (i.e.\ the spatial distribution) of people in a given area at various instants of time. 
By computing a suitable approximation of the Wasserstein distance between two consecutive density profiles, we are able to extract the main directions followed by people, i.e.\ to understand how the mass of people distribute in space and time.
The main applications of the proposed technique are the monitoring of daily flows of commuters, the organization of large events, and, more in general, the traffic management and control.
\end{abstract}


\end{frontmatter}

\section{Introduction}
Since many years researchers use data from cellular networks to extrapolate useful information about social dynamics. The interested reader can find in the survey paper \cite{blondel2015SPRINGER} an exhaustive list of possible uses of such a data.
The main reason for this large interest lies in the fact that, nowadays, basically all of the people in the developed world own a mobile phone (with or without internet connection). Therefore, we can get a complete view of the positions of people considering the location of the fixed antennas each device is connected to. Moreover, the huge amount of available data counterbalances in part the fact that device positioning techniques generally provide poor spatial and temporal accuracy (much less than the GPS, for example). 

In this paper we are interested in models and methods to inferring activity-based human mobility flows from mobile phone data. Among papers which investigate the usage of mobile phone data in this direction, many of them involve Call Detail Records or similar types of data, see, e.g., \cite{becker2013ACM, iqbal2014ELSEVIER, jarv2014ELSEVIER, gonzalez2008N, jiang2017IEEE, naboulsi2013, zheng2016IEEE}.
Other papers use aggregated data such as those coming from Erlang measurements, see, e.g., \cite{calabrese2011IEEE, reades2009SAGE, sevtsuk2010TF}.

In this paper, instead, data consist of density profiles (i.e.\ the spatial distribution) of people in a given area at various instants of time. Mobile devices are not singularly tracked, but their logs are aggregated in order to obtain the total number of users in a given area. Such a data, not publicly available at the moment, are provided by the Italian telecommunication company TIM. 

The goal of the paper is to ``assign a direction'' to the presence data.
In fact, the mere representation of time-varying density of people clearly differentiate attractive from repulsive or neutral areas but does not give any information about the directions of flows of people. In other words, we are interested in a ``where-from-where-to'' type of information, which reveals travel flows and patterns of people. 
The goal is pursued by computing a suitable approximation of the Wasserstein distance (also known as `earth mover's distance' or `Mallows distance') between two consecutive density profiles. The computation of the Wasserstein distance gives, as a by-product, the \emph{optimal flow} which, in our case, coarsely corresponds to the main directions followed by people, i.e.\ how the mass of people distribute in space and time. 
It is useful to note here that the same methodology is investigated in the recent paper \cite{zhu2018IJGIS}, where similar phone data are used and similar results are obtained.\footnote{\cite{zhu2018IJGIS} was published after the  submission of this paper and we have been aware of it during the review process.}

The applicability of this approach is \emph{a priori} questionable since it is based on many assumptions that are, in general, very far to be true. Let us mention here the fact that people can move in any direction of the space neglecting hard obstacles and that they are indistinguishable and interchangeable. Moreover, we are not able to distinguish vehicular from pedestrian traffic.\\
In spite of this strong assumptions, numerical simulations presented here show that our approach leads to very meaningful results, and then it can be actually employed in traffic management and control. We think that the main applications of the technique proposed here can be the monitoring of daily flows of commuters and the organization of large events.

\section{Dataset}\label{sec:data}
TIM provides estimates of mobile phones presence in a given area in raster form: the area under analysis is split into a number of elementary territory units (ETUs) of the same size (about $150\times 150$ m$^2$ in urban areas). The estimation algorithm does not singularly recognize users and does not track them using GPS. It simply counts the number of phone attached to network nodes and, knowing the location and radio coverage of the nodes, estimates the number of TIM users within each ETU at any time. 
TIM has now a market share of 30\% with about 29.4 million mobile lines in Italy (AGCOM, Osservatorio sulle comunicazioni 2/2017).

\begin{figure}[h!]
\begin{center}
\includegraphics[width=8.5cm]{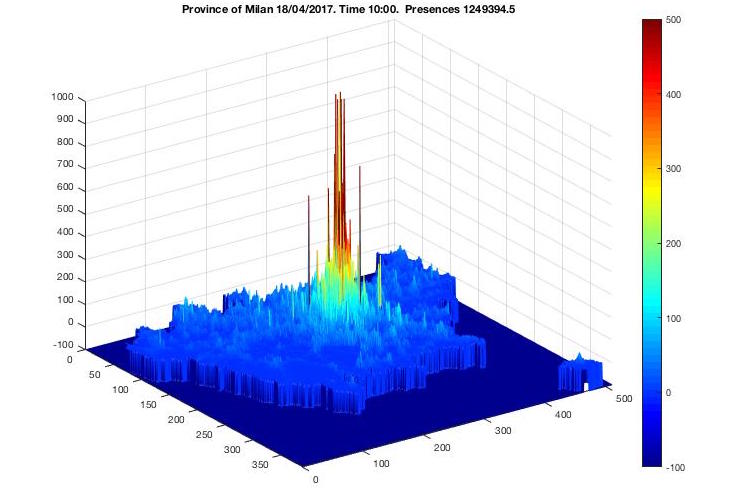}    
\caption{3D-plot of the number of TIM users in each ETU of Milan's province on April 18, 2017.} 
\label{fig:presenze3D}
\end{center}
\end{figure}

\begin{figure}[h!]
\begin{center}
\includegraphics[width=8.7cm,height=5.2cm]{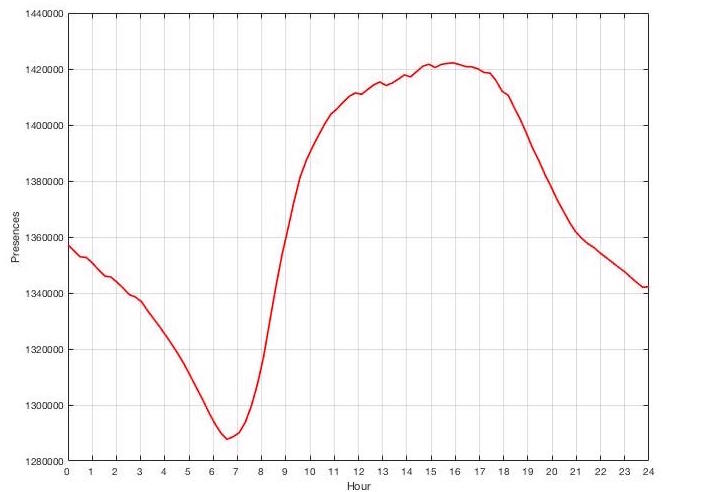}    
\caption{Trend of presences in the province of Milan during a typical working day.} 
\label{fig:presenzeGiorno}
\end{center}
\end{figure}

\begin{figure}[h!]
\begin{center}
\includegraphics[width=8.7cm,height=5.2cm]{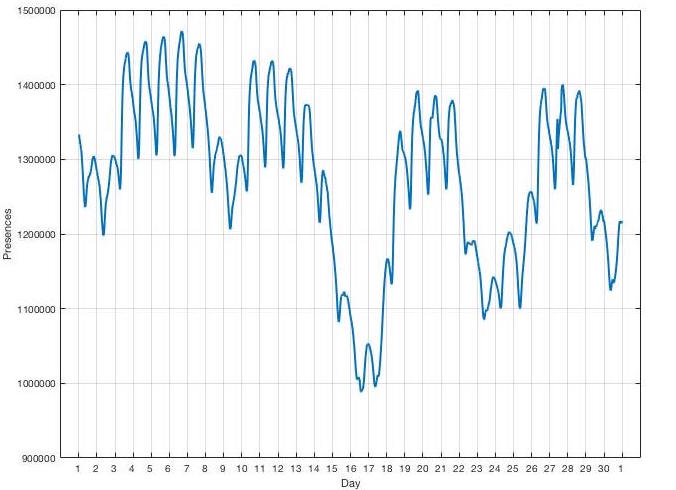}  
\caption{Trend of presences in the province of Milan during April 2017.} 
\label{fig:presenzeMese}
\end{center}
\end{figure}

The data we worked with refer to the area of the province of Milan (Italy), which is divided in 198,779 ETUs, distributed in a rectangular grid $511\times 389$. Data span six months (February, March and April 2016 and 2017). The entire period is divided into time intervals of 15 minutes, therefore we have 96 data per day per ETU in total. \\ 
In Fig.~\ref{fig:presenze3D} we graphically represent presence data at a fixed time. We observe that the peak of presence is located in correspondence of Milan city area.
Fig.~\ref{fig:presenzeGiorno} shows the presences in the province of Milan in a typical working day. The curve in the image decreases during the night, it increases in the day-time and decreases again in the evening. These variations are due to two main reasons: first, the arrival to and departure from Milan's province of visitors and commuters. Second, the fact that when a mobile phone is switched off or is not communicating for more than six hours, its localization is lost. 
The presence value that most represents the population of the province is observed around 9  pm., when an equilibrium between traveling and phone usage is reached. This value changes between working days and weekends, but it is always in the order of $1.3\times10^6$.
Fig.~\ref{fig:presenzeMese} shows the trend of presence data during April 2017. We can observe a cyclical behavior: in the working days the number of presences in the province is significantly higher than during the weekends. It is interesting to note the presence of two low-density periods on April 15-18 and on April 22-26, 2017, determined respectively by the Easter and the long weekend for the Italy's Liberation Day holiday.

\section{Mathematical model}
Our purpose is to analyze the flow of people from presence data. To do that, let us first introduce the Monge--Kantorovich mass transfer problem, see~\cite{villani2008SSBM}, which can be easily explained as follows: 
given a sandpile with mass distribution $\rho_0$ and a pit with equal volume and mass distribution $\rho_1$, find a way to minimize the cost of transporting sand into the pit. The cost for moving mass depends on both the distance from the point of origin to the point of arrival and the amount of mass it is moved along that path. We are interested in minimizing this cost by finding the optimal paths to transport the mass from the initial to the final configuration.

This approach goes through the notion of \emph{Wasserstein distance}, see again~\cite{villani2008SSBM}.
%
%
In the space $\R^n$ equipped with the euclidean metrics, let $\rho^{0}$ and $\rho^{1}$ be two density functions such that $\int_{\R^n}\rho^0=\int_{\R^n}\rho^1$. For all $p\in[1,+\infty)$, the $L^p$-Wasserstein distance between $\rho^0$ and $\rho^1$ is
\begin{equation}\label{WassRn}
W_p(\rho^0,\rho^1)=\bigg(\min_{T\in\mathcal{T}}\int_{\R^n}\|T(x)-x\|_{\R^n}^p \, \rho^0(x)dx\bigg)^{\frac{1}{p}}
\end{equation}
where
\begin{align*}
\mathcal{T}:=\Biggr\{T\colon\R^n\to\R^n \, : \, \int\displaylimits_B &  \rho^1(x)dx =\int\displaylimits_{\{x:T(x)\in B\}} \rho^0(x)dx, 
\\ \forall \, B\subset\R^n \text{ bounded}\Biggr\}.
\end{align*}
$\mathcal{T}$ is the set of all possible maps which transfer the mass from one configuration to the other. 
The physical interpretation of this definition is naturally related to the solution of the Monge--Kantorovic problem since Wasserstein distance corresponds to the minimal cost needed to rearrange the initial distribution $\rho^0$ into the final distribution $\rho^1$.
\begin{remark} We are not interested in the actual value of the Wasserstein distance $W_p$, instead we look for the \emph{optimal map} $T^*$ which realizes the $\arg\min$ in \eqref{WassRn}, and represents the paths along which the mass is transferred.
\end{remark}

\medskip

Following \cite{briani2017CMS}, we reformulate the mass transfer problem on a graph $\mathcal{G}$ with $N$ nodes. This procedure gives an approximation of the Wasserstein distance \eqref{WassRn} and provides an algorithm for computing optimal paths.
Starting from an initial mass $m^0_j$ and a final mass $m^1_j$, for $j=1,\dots,N$, distributed on the graph nodes, we aim at rearranging in an optimal manner the first mass in the second one. We denote by  $c_{jk}$ the cost to transfer a unit mass from node $j$ to node $k$, and by $x_{jk}$ the (unknown) mass moving from node $j$ to node $k$. The problem is then formulated as
\[ 
\text{minimize }\mathcal{H}:= \sum_{j,k=1}^N c_{jk}x_{jk}
\]
subject to
\[\displaystyle\sum_k x_{jk}=m_j^0 \,\,\,\, \forall j, \quad
\displaystyle\sum_j x_{jk}=m_k^1 \,\,\,\, \forall k \quad\text{and}\quad x_{jk}\geq 0.\]
Defining
\begin{flalign*}
&x = (x_{11}, x_{12},\dots,x_{1N},x_{21},\dots,x_{2N},\dots,x_{N1},\dots,x_{NN})^T,\\
&c= (c_{11}, c_{12},\dots,c_{1N},c_{21},\dots,c_{2N},\dots,c_{N1},\dots,c_{NN})^T,\\
&b = (m^0_1,\dots,m^0_N,m^1_1,\dots,m^1_N)^T,
\end{flalign*}
and the matrix
\[A =\begin{bmatrix}
\bbone_N & 0 & 0 & \dots & 0\\[0.3em]
0 & \bbone_N & 0 & \dots & 0\\[0.3em]
0 & 0 & \bbone_N &  \dots & 0\\[0.3em]
\vdots & \vdots & \vdots & \ddots & \vdots\\[0.3em]
0 & 0 & 0 & \dots & \bbone_N\\[0.3em]
I_N & I_N & I_N & I_N & I_N\\[0.3em]
\end{bmatrix},\]
where $I_N$ is the $N\times N$ identity matrix and $\bbone_N=\displaystyle\underbrace{(1\  1 \dots \ 1)}_{N\text{ times}}$, our problem is written as a standard linear programming (LP) problem: minimizes $\displaystyle c^Tx$, under the conditions $Ax=b$ and $x\geq 0$, see \cite[Sec.\ 6.4.1]{santambrogio2015SPRINGER} and \cite[Chap.\ 19]{sinha2005ELSEVIER}.
The result of the algorithm is a vector $x^*:=\arg\min c^T x$ whose elements $x^*_{jk}$ represent how much mass moves from node $j$ to node $k$ employing the minimum-cost mass rearrangement. 

\section{Application to human mobility flows}
In this paragraph we describe the application of the LP-based mass transfer problem to TIM data. 
First of all, we exploit the subdivision into ETUs of the province of Milan (see Section \ref{sec:data}), considering a graph whose nodes coincide with the centers of such ETUs. We assume that each node is connected to all the others. 
Therefore, the amount of people located in each ETU $j$ represents the mass $m_j$ to be moved. 
Solving the LP problem with two consecutive (in time) mass distributions $m^0$ and $m^1$, we get the optimal path followed by people to move from the first configuration to the second one. 

Now we focus on the definition of the cost function $c$. This function is related to the distance between the starting point and the arrival point, so it would make sense to use the standard Euclidean distance.  On the other hand, this choice can lead to nonphysical optimal displacements, as we can see in the following example.

\begin{example}
We have to move one to the right three unit masses, using the Euclidean distance as cost function. In the first scenario (see Fig.~\ref{fig:percorso1}) all masses move one to the right, while in the second scenario (see Fig.~\ref{fig:percorso2}) the leftmost mass move three to the right and the other two are frozen. Although the two mass movements are different, the Wasserstein distance is the same and equal to three.
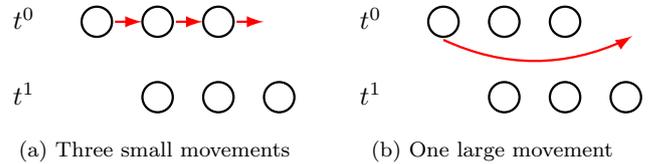
\begin{figure}[h!] 
\centering
\setlength{\unitlength}{0.4cm}
\subfloat[][Three small movements]
{\begin{picture}(9.5,4)(0,0)
\linethickness{0.3mm}
\put(0.25,3.25){\makebox{$t^0$}}
\put(3,3.5){\circle{1}}\color{red}
\put(3.6,3.5){\vector(0.9,0){0.9}}\color{black}
\put(5,3.5){\circle{1}}\color{red}
\put(5.6,3.5){\vector(0.9,0){0.9}}\color{black}
\put(7,3.5){\circle{1}}\color{red}
\put(7.6,3.5){\vector(0.9,0){0.9}}
\color{black}
\put(0.25,0.75){\makebox{$t^1$}}
\put(5,1){\circle{1}}
\put(7,1){\circle{1}}
\put(9,1){\circle{1}}
\end{picture} 
\label{fig:percorso1}}\quad\,
\subfloat[][One large movement]
{\begin{picture}(9.5,4)(0,0)
\linethickness{0.3mm}
\color{black}
\put(0.25,3.25){\makebox{$t^0$}}
\put(3,3.5){\circle{1}}
\put(5,3.5){\circle{1}}
\put(7,3.5){\circle{1}}
\color{red}
\qbezier(3,2.9)(6,1.5)(9,2.9)
\put(9.1,2.96){\vector(0.08,0.05){0.12}}
\color{black}
\put(0.25,0.75){\makebox{$t^1$}}
\put(5,1){\circle{1}}
\put(7,1){\circle{1}}
\put(9,1){\circle{1}}
\end{picture}
\label{fig:percorso2}}
\caption{Different mass movements with equal Wasserstein distance.}
\label{fig:WassEquivalenti}
\end{figure} 
\end{example}

Small movements seem to better describe the flow of large crowds. Therefore, in order to select small movements rather than large ones, we slightly modify the cost function as follows:
\begin{equation}
c(P,Q)=\|P-Q\|_{\mathbb R^2}^{1+\varepsilon},
\label{eq:costo}
\end{equation}
where $P$ and $Q$ are the centers of two ETUs (nodes of the graph) and $\varepsilon>0$ is a small parameter. By means of parameter $\varepsilon$ (0.1 in our simulations) we increase the cost of large movements in favor of small ones.

\begin{remark}
Recalling the definition of Wasserstein distance,
the mass flowing along the graph must be preserved in time, i.e.\ $\sum_{j} m^0_j=\sum_{j} m^1_j$. 
The data we work with do not strictly verify this property, so we have modified the mass in two different ways:
1. distributing the excess mass along the boundary of the considered area; 
2. distributing the excess mass uniformly in the considered area. 
In both cases the mass modification allows the algorithm to be correctly implemented but, by analyzing the results, we have found that it is better to proceed by distributing the excess mass uniformly.
\end{remark}

As already mentioned in the Introduction, people's behavior does not match the assumptions on which the optimal mass transfer problem is originally built. Beside the fact that people cannot freely move in the space, in general the crowd does not move in such a way to minimize the total displacement as a whole (even if a sort of ``minimal-effort'' assumption could be realistic for single persons).
In the next section we will see that these deviations from constitutive assumptions seem to be, at least to some extent, negligible.


\section{Numerical results}
The LP problem is solved using as inputs all the pairs $(m^0,m^1)$ corresponding to the number of people at two consecutive time instants for the whole day (95 LP problems in total each day). 
We denote by $(x^*)^n$, $n=0,\ldots,94$ the solution of the LP problem between time instants $t^n$ and $t^{n+1}$, where $t^n=00:00+n\cdot 15$min.
Only movements larger than the daily average $M$ are drawn, with $M$ defined as
\[M:=\displaystyle\frac{1}{N_{nz}} \sum_{n=0}^{94} \ \sum_{\substack{j,k=1\\j\neq k}}^{N}(x^*)_{jk}^n,\]
where $N_{nz}$ is the number of non-zero values. 
%
Note that for $j=k$, the value $x^*_{jj}$ gives the mass which remains in the ETU $j$ between the two times. The following example explains why we exclude them from the set of significant movements.
\begin{example}
Let us consider the graph with two nodes shown in Fig.~\ref{fig:esempioFrecce}, which has mass 12 in node 1 and mass 16 in node 2 at time $t^0$. We assume that in the time interval between $t^0$ and $t^1$ a mass equal to 8 is moved from node 1 to node 2 and a mass equal to 10 is moved from node 2 to node 1. 
At time $t^1$ we have both the node 1 and 2 with mass 14. The vector $x$ which describes the flow of mass is
\[ x_{11}=4 \quad\,\,\, x_{12}=8 \quad\,\,\, x_{21}=10 \quad\,\,\, x_{22}=6,\]
while, the LP algorithm gives as a solution
\[x^*_{11}=12 \quad\,\, x^*_{12}=0 \quad\,\,\, x^*_{21}=2 \quad\,\,\,\, x^*_{22}=14.\]
This is because the algorithm has only information about initial and final mass distribution and solves a minimum problem. Therefore, since the cost of a null shift is certainly preferable to any other movement, elements $x^*_{jj}$ generally have large values, but they do not represent a real mass transfer and are not significant for the flow analysis. 
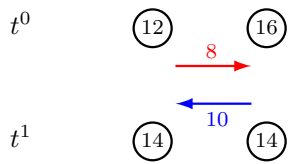
\begin{figure}[h!]
\begin{center}
\setlength{\unitlength}{0.5cm}
\begin{picture}(15,4)(0,0)
\linethickness{0.3mm}
\color{black}
\put(2.25,3.36){\makebox{$t^0$}}
\put(6,3.5){\circle{1}}
\put(5.7,3.35){\makebox{\small 12}}
\put(9,3.5){\circle{1}}
\put(8.7,3.35){\makebox{\small 16}}
\color{red}
\put(6.6,2.5){\vector(2,0){2}}
\put(7.4,2.7){\makebox{\small 8}}
\color{blue}
\put(8.6,1.5){\vector(-2,0){2}}
\put(7.4,0.9){\makebox{\small 10}}
\color{black}
\put(2.25,0.36){\makebox{$t^1$}}
\put(6,0.5){\circle{1}}
\put(5.7,0.35){\makebox{\small 14}}
\put(9,0.5){\circle{1}}
\put(8.7,0.35){\makebox{\small 14}}
\end{picture}
\caption{Example of mass movements in a graph with two nodes between time $t^0$ and $t^1$.} 
\label{fig:esempioFrecce}
\end{center}
\end{figure}
\end{example}
 
In the following figures flows will be represented by arrows that join departure and arrival ETUs. The gray level of the arrows depends on the intensity of the flow, i.e.\ the amount of people actually moving. 
%
%
To implement the algorithm we have used Matlab, in particular its function $\texttt{linprog}$ for solving the LP problems. 

Finally, note that we are not able to analyze the whole area of the province of Milan. This is because, considering the whole graph, the matrix $A$ would have size $2N\times N^2\sim1.5 \times 10^{16}$ and would be unmanageable for both the amount of memory required and the computing times. For this reason, we either analyzed smaller areas, focusing on the most significant ones, or we considered large areas aggregating data of neighboring ETUs.

\subsection{Test 1. Macroscopic scale: flows of commuters}
\begin{figure}[h!]
\begin{center}
\includegraphics[width=8.4cm,height=6cm]{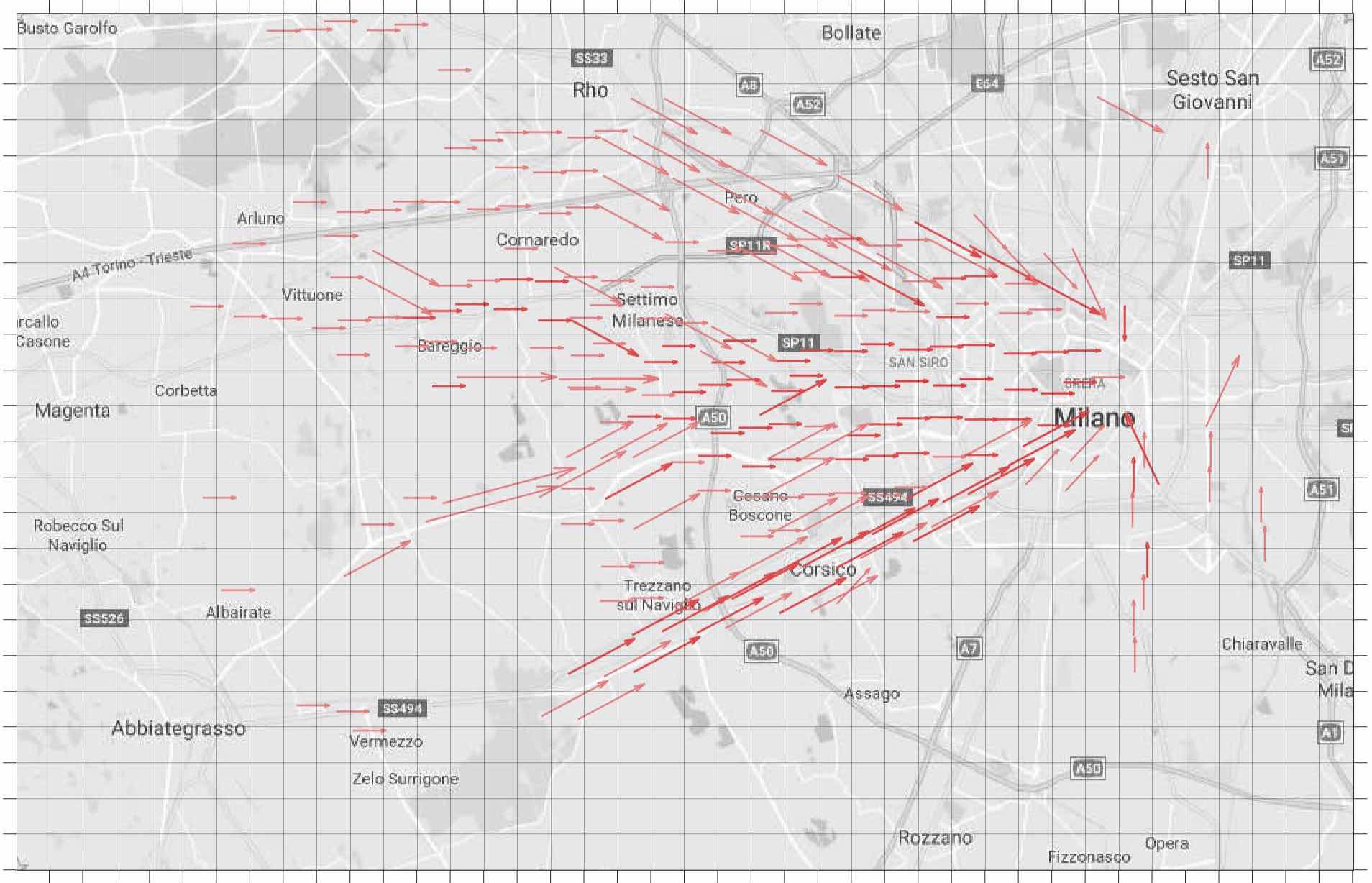}   
\caption{Test 1: main flows around Milan's area during a generic working day in the morning.} 
\label{fig:test1M}
\end{center}
\end{figure}

\begin{figure}[h!]
\begin{center}
\includegraphics[width=8.4cm,height=6cm]{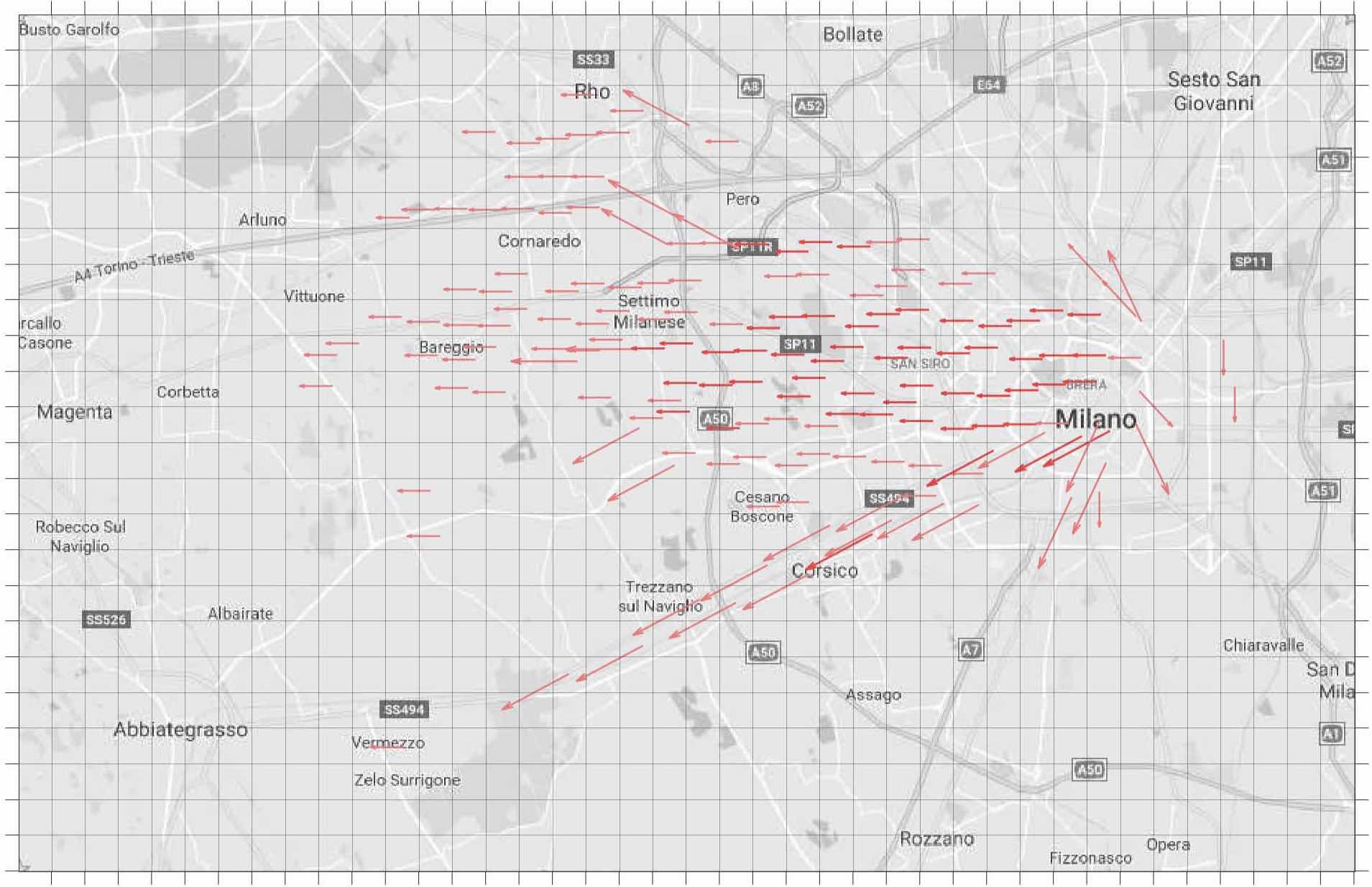}   
\caption{Test 1: main flows around Milan's area during a generic working day in the evening.} 
\label{fig:test1S}
\end{center}
\end{figure}

The area shown in Figs.~\ref{fig:test1M}-\ref{fig:test1S} is a rectangle $40\times 24$ contained in the province of Milan that has been obtained by aggregating ETUs into groups of $6\times 6$. The pictures show the main flows on a generic working day in the morning and in the evening. It is clear that the flows are directed towards and from the city of Milan, and are mainly determined by commuters. In particular we can see movements from/to the left of the province to Milan.

\subsection{Test 2. Aggregated flow along main roads}
\begin{figure}[h!]
\begin{center}
\includegraphics[height=2.1cm]{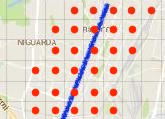}   
\caption{Test 2: ETUs around one of the selected main roads whose flows are aggregated and gathered along the road.} 
\label{fig:test2CV}
\end{center}
\end{figure}

We have chosen 8 main roads which lead to the city of Milan in order to visualize only the flows along some predefined directions. To this end, we have localized the ETUs in a neighborhood  of the roads and we summed all the flows pointing from these ETUs to the others in the neighborhood. Finally, we have aggregated the resulting flow along the considered roads, see Fig.~\ref{fig:test2CV}.

\begin{figure}[h!]
\begin{center}
\includegraphics[width=8.4cm,height=6cm]{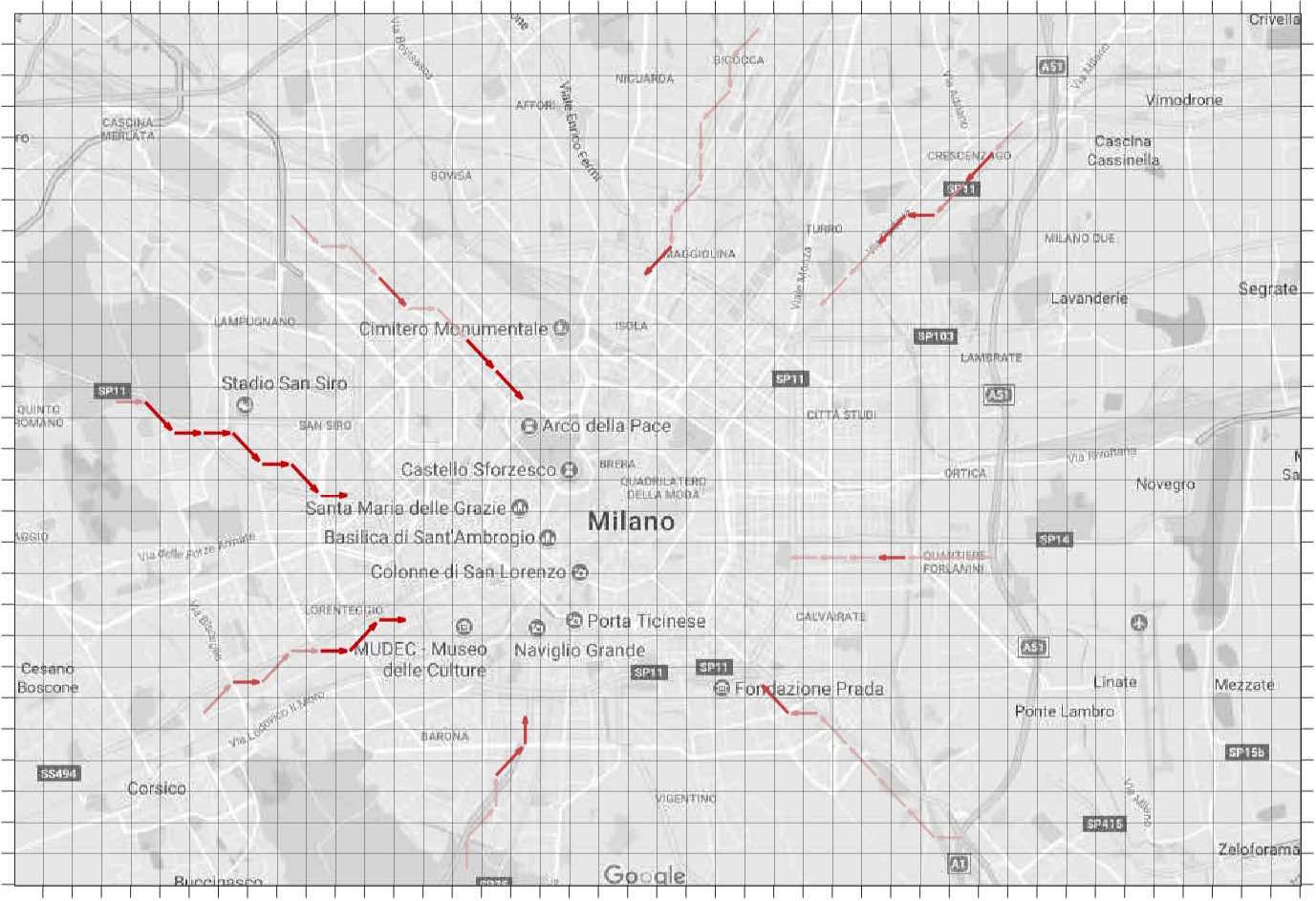}   
\caption{Test 2: flows in Milan's area along selected main roads during a generic working day in the morning.} 
\label{fig:test2M}
\end{center}
\end{figure}

\begin{figure}[h!]
\begin{center}
\includegraphics[width=8.4cm,height=6cm]{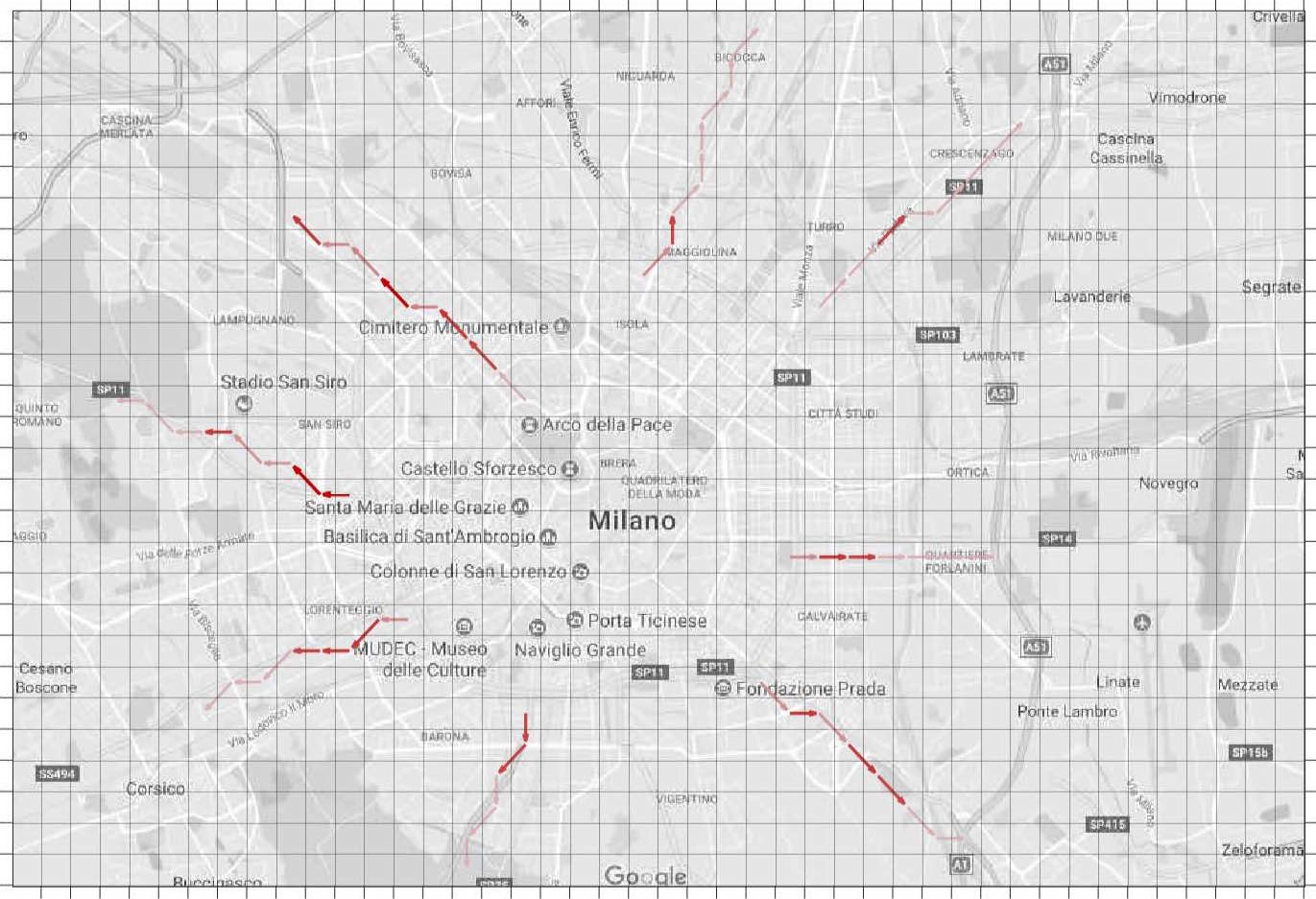}   
\caption{Test 2: flows in Milan's area along selected main roads during a generic working day in the evening.} 
\label{fig:test2S}
\end{center}
\end{figure}

The result obtained by this process can be seen in Figs.~\ref{fig:test2M}-\ref{fig:test2S}. The considered area is a rectangle $44\times 28$ contained in the province of Milan that has been obtained by aggregating ETUs into groups of $3\times 3$. The pictures show the main flows located at the eight specifically defined directions on a generic working day between 8:15 and 8:30 am and between 5:45 and 6:00 pm.
By observing the images we can easily identify the main directions of the flow and the roads with more traffic load.


\subsection{Test 3. Flows influenced by a large event}
\begin{figure}[h!]
\begin{center}
\includegraphics[width=0.3\textwidth]{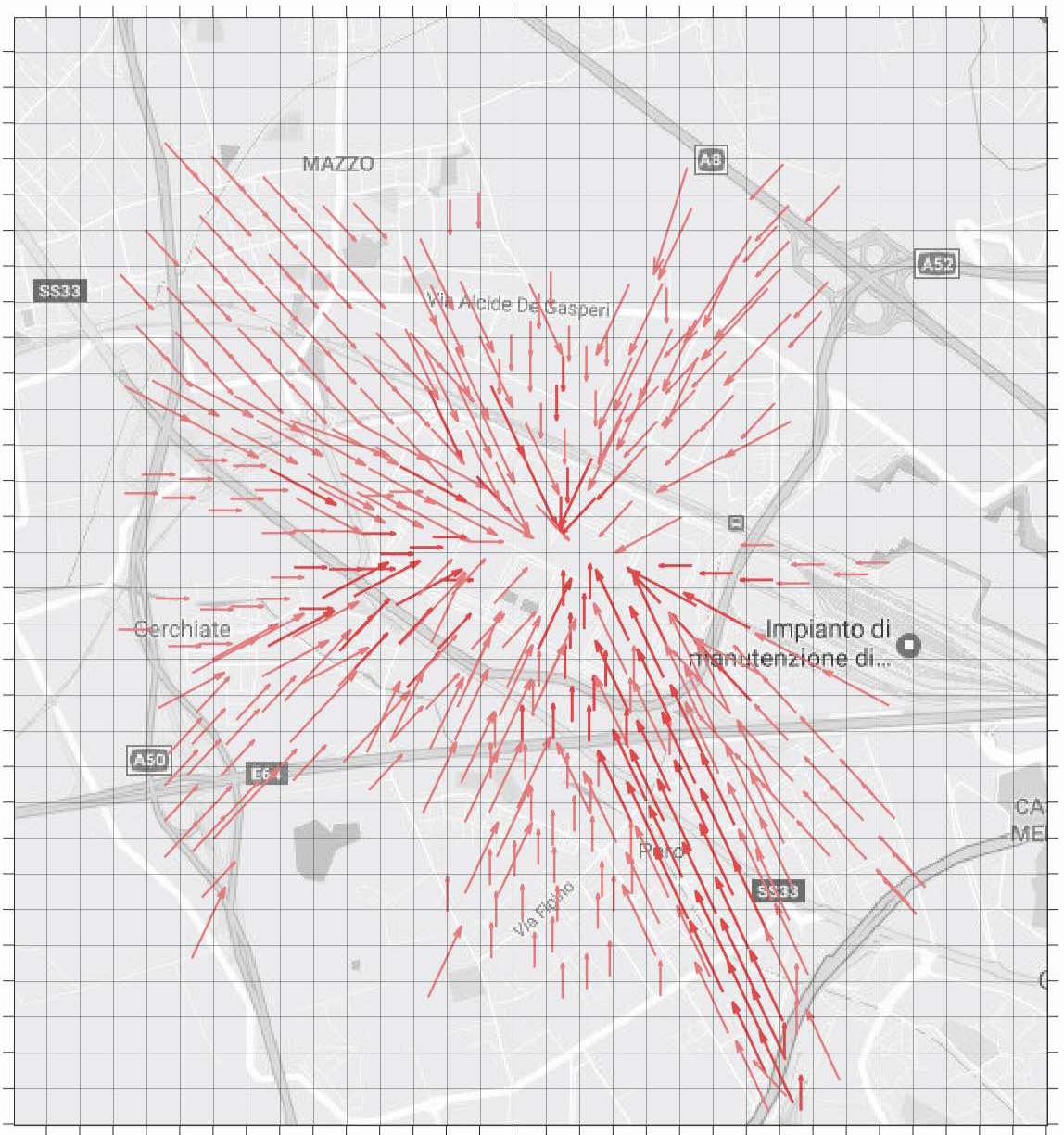} 
\caption{Test 3: flows directed to the area of the exhibition between 9:45 and 10:00 am.} 
\label{fig:test3Mattina}
\end{center}
\end{figure}

\begin{figure}[h!]
\begin{center}
\includegraphics[width=0.3\textwidth]{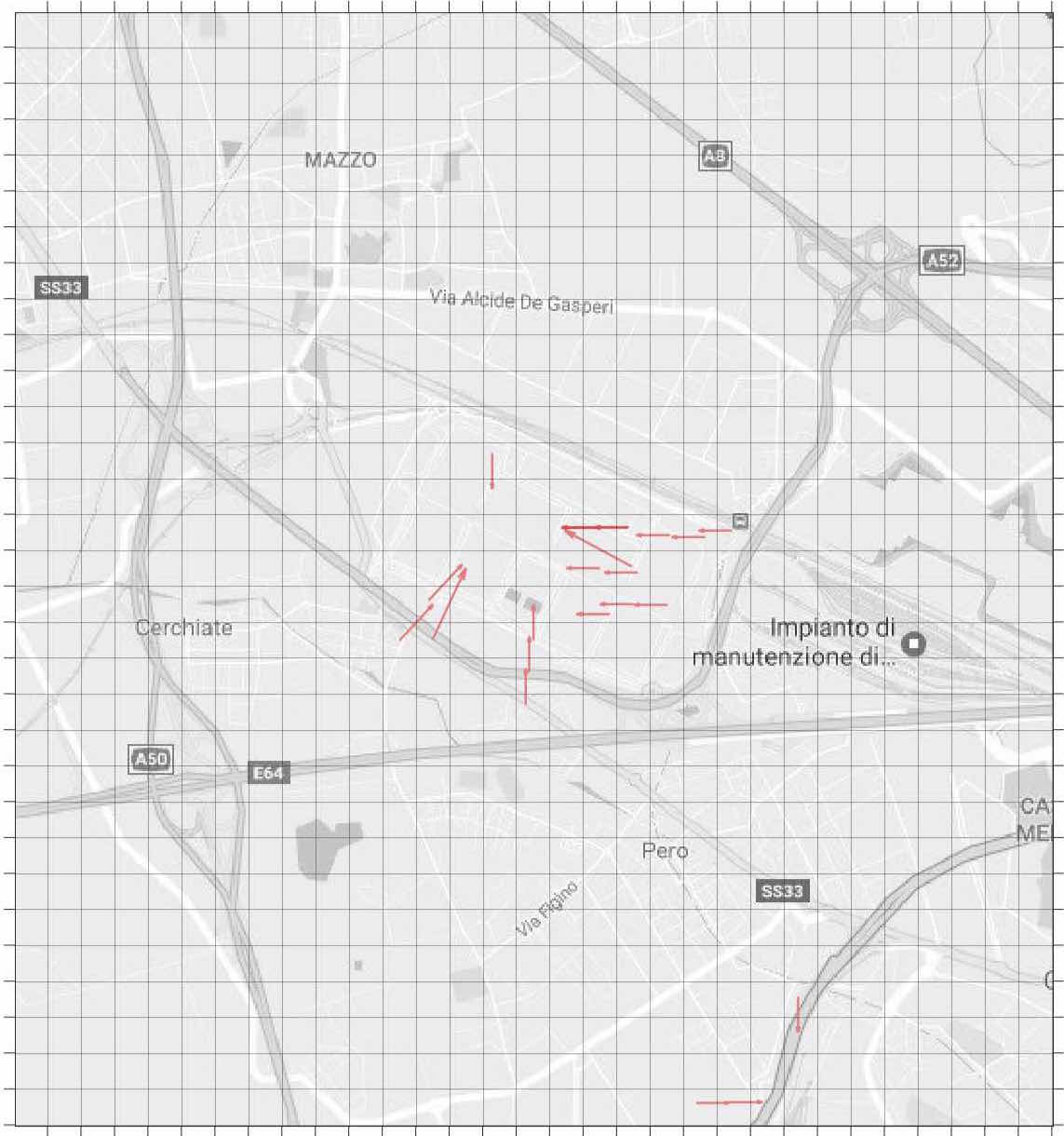}    
\caption{Test 3: flows at the area of the exhibition between 1:00 and 1:15 pm.} 
\label{fig:test3Pranzo}
\end{center}
\end{figure}

\begin{figure}[h!]
\begin{center}
\includegraphics[width=0.3\textwidth]{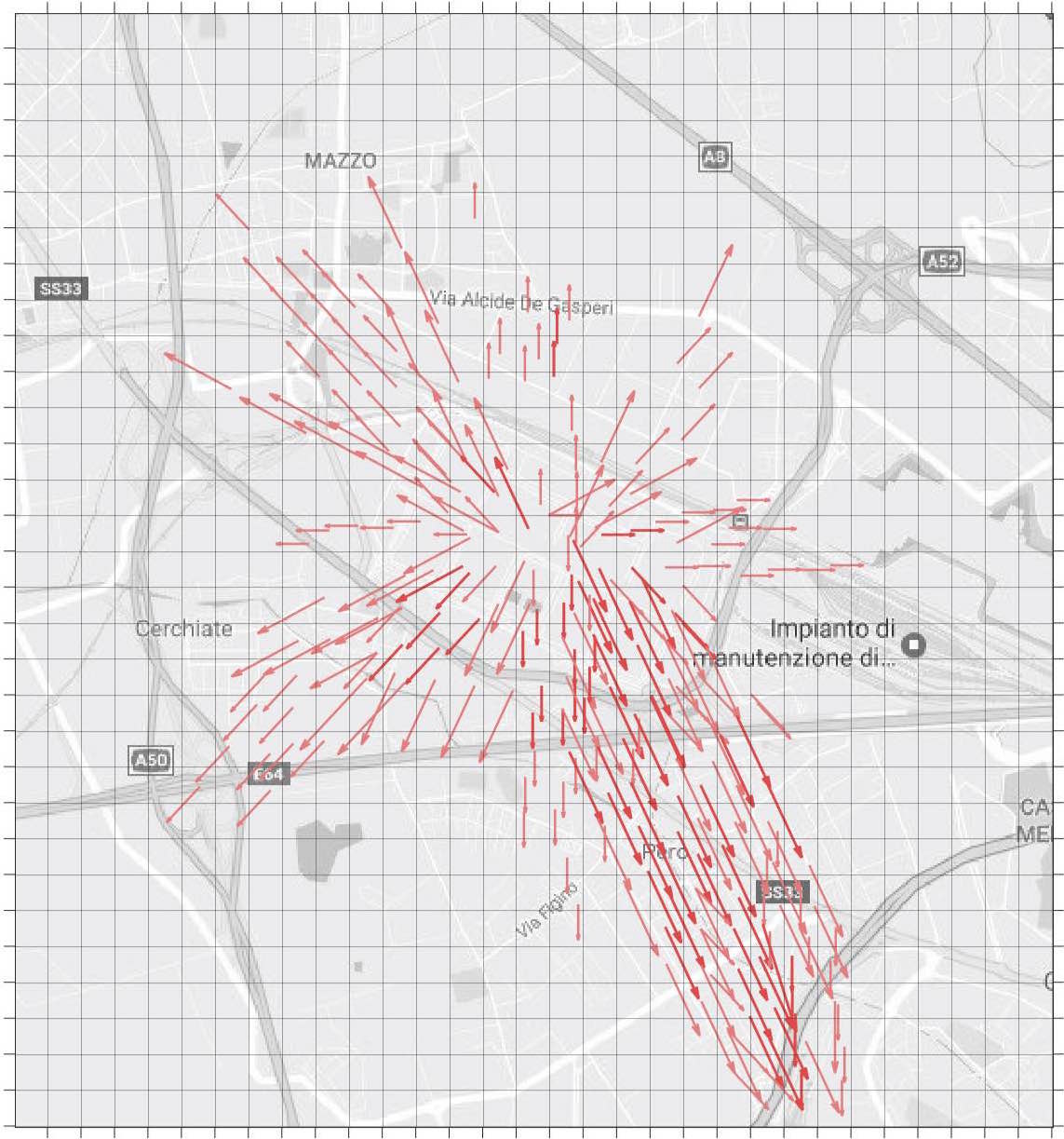}   
\caption{Test 3: flows leaving the area of the exhibition between 5:45 and 6:00 pm.} 
\label{fig:test3Sera}
\end{center}
\end{figure}

In this test we show the effects of a large event on urban mobility. The event we have analyzed is the exhibition  of the \emph{Salone del Mobile}, held every April at Fiera Milano exhibition center in Rho, near Milan. The area in Figs.~\ref{fig:test3Mattina}-\ref{fig:test3Sera} is a square $31\times 31$ contained in Rho area and centered around Fiera Milano. We show three different behavior of flows during the exhibition.
Fig.~\ref{fig:test3Mattina} shows the main flows to Fiera Milano at the opening of the exhibition in the morning. We can observe that the more significant arrows are directed to the exhibition. 
Fig.~\ref{fig:test3Pranzo} shows the main flows during lunch time. In this case we find very few arrows because there are no really significant movements and no preferred directions. 
Finally, Fig.~\ref{fig:test3Sera} shows a similar behavior to the morning time, with a reverse direction of the flow, due to the closure of the exhibition. 
It is interesting to note that both in the morning and in the evening, the most intense flows are in the South East part of the map, in correspondence of the roads that join the city of Milan with Fiera Milano.

\subsection{Test 4. Microscopic scale: accesses to the exhibition}
\begin{figure}[h!]
\begin{center}
\includegraphics[width=8.5cm,height=6.5cm]{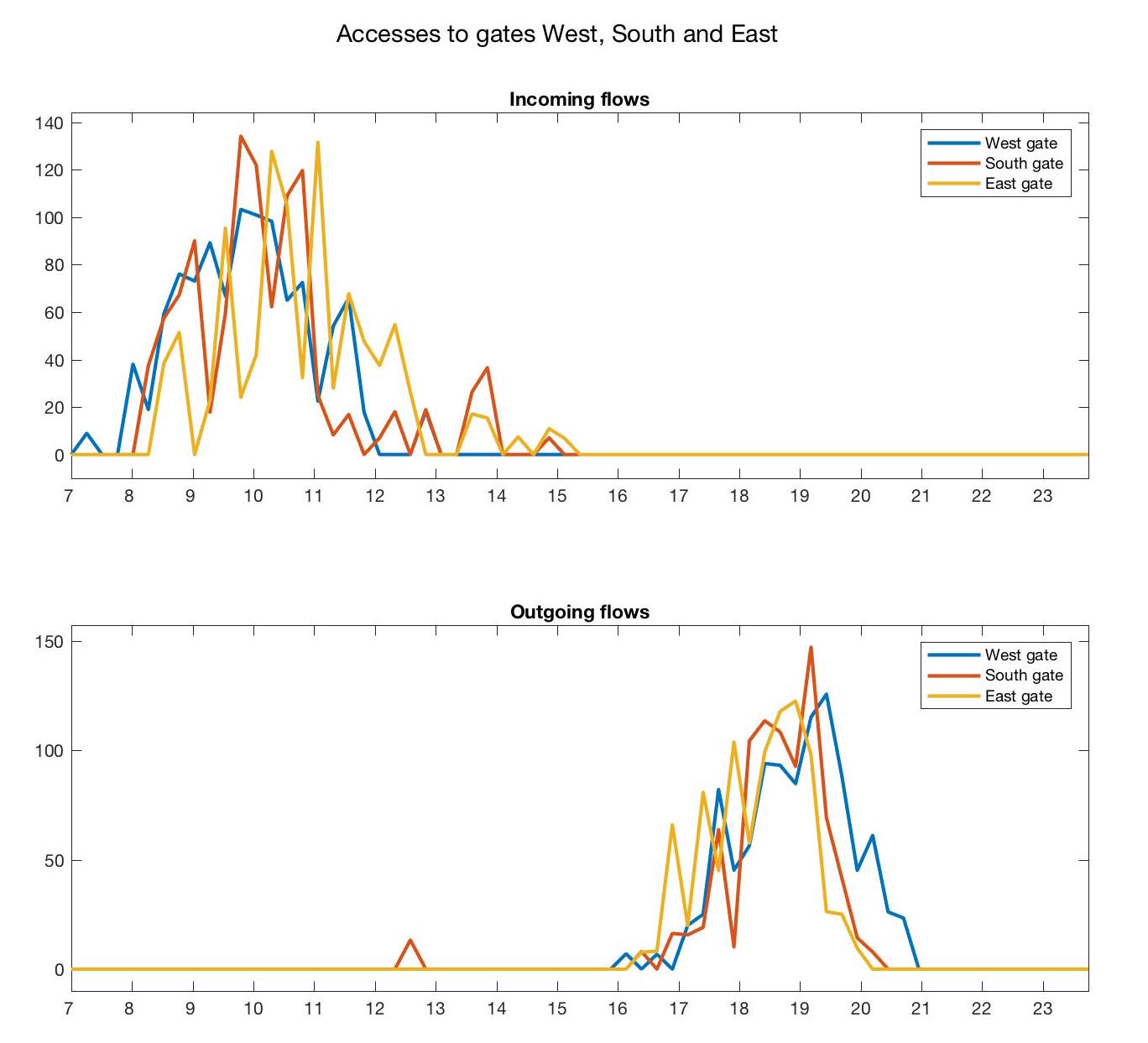}   
\caption{Test 4: incoming and outgoing flows from the West, South and East gates of Fiera Milano on the first day of the exhibition.} 
\label{fig:test4}
\end{center}
\end{figure}

In this last test, we consider a very small area to catch the flows to/from the ETUs corresponding to access points at Fiera Milano. We show the first day of the exhibition of the Salone del Mobile. We define incoming and outgoing flows as follows: the incoming flows are given by the sum of the flows from the outside of the exhibition to the gates and the flows from the gates to the inside of the exhibition; the outgoing flows are given by the sum of the flows from the inside of the exhibition to the gates and the flows from the gates to the outside of the exhibition.
Fig.~\ref{fig:test4} shows the incoming and the outgoing flows as a function of time during the whole day. By looking at the plots we can identify which gate is the most used by the visitors.

\section{Conclusions}
This paper aimed at understanding how the mass of people distribute on large areas by a coarse estimation of their locations at consecutive snapshots. Despite the strong constitutive assumptions, the Wasserstein distance allows to get useful information and deserves further investigations.
Future work will aim at applying this approach to construct O-D matrices from the optimal map and to control and estimate traffic states. 
Comparisons with other techniques and the link to different types of transportations metrics will be also investigated.
At the same time, more performing implementations will be considered and analysed.

\bibliography{ifacconf}  
\end{document}